\documentclass[a4paper]{jpconf}

\usepackage{graphicx}

\begin{document}

\def\arcsec{\hbox{$^{\prime\prime}$}}
\def\farcs{\hbox{$.\!\!^{\prime\prime}$}}
\def\degr{\hbox{$^\circ$}}

\title{AstraLux - the Calar Alto 2.2-m telescope Lucky Imaging camera}
\author{F. Hormuth, W. Brandner, S. Hippler, Th. Henning}
\address{Max-Planck-Institute for Astronomy, K\"onigstuhl 17, 69117 Heidelberg, Germany}
\ead{hormuth@mpia.de}

\begin{abstract}
AstraLux is a Lucky Imaging camera for the Calar Alto 2.2-m telescope, based on an 
electron-multiplying high speed CCD. 
By selecting only the best 1-10\% of several thousand short exposure frames, AstraLux 
provides nearly diffraction limited imaging capabilities in the SDSS {\em i'} and {\em z'}\ filters 
over a field of view of 24$\times$24 arcseconds. \\
By choosing commercially available components wherever possible, the instrument could 
be built in short time and at comparably low cost. We briefly present the instrument design, the data 
reduction pipeline, and summarise the performance and characteristics\footnote{Based on 
observations collected at the Centro Astron\'omico Hispano Alem\'an (CAHA) at Calar Alto, 
operated jointly by the Max-Planck Institut f\"ur Astronomie and the Instituto de Astrof\'isica 
de Andaluc\'ia (CSIC).}.
\end{abstract}


\section{Introduction}
The recovery of the full theoretical resolution of large ground-based optical telescopes
in the presence of atmospheric turbulence has been a major goal of technological
developments in the field of astronomical instrumentation in the past decades. 
The de-facto standard nowadays is adaptive optics, the active real-time correction of the incoming 
distorted wavefronts by means of deformable mirrors. While diffraction limited 
image quality can be routinely achieved in the near infrared at large telescopes, this is 
only possible with considerable technical effort.

Besides speckle techniques in the Fourier domain, a relatively simple 
approach is the selection of images based on their quality: instead of using all images
of a large set of short-time exposures, only those showing little image distortion due to the
variable strength of atmospheric turbulence are
combined to a high-resolution result. Like in the speckle interferometric approach,
this requires only a detector with fast readout capability and moderate computational
effort. 
The application of this technique to fainter astronomical targets, e.g. low-mass double stars, was limited
by the readout noise of the available detectors at visible wavelengths until a few years ago. At typical exposure times of few ten
milliseconds -- necessary to ``freeze'' the effects of atmospheric turbulence -- photon noise
limited detectors are obligatory to use this so-called ``Lucky Imaging'' technique on targets fainter than $\approx$10\,mag. 
Since the image quality of each single frame has to be determined, e.g. by measuring the Strehl ratio of a 
reference star, the readout noise of the detector sets stringent limits on the minimum brightness of this reference.

The advent of electron multiplying CCDs (EMCCD) with single-photon detection capabilities led
to considerable interest in the Lucky Imaging technique on the part of professional astronomers \cite{Jerram:2001,Hynecek:2002,Mackay:2001}. 
First experiments at the Nordic Optical Telescope (NOT) with LuckyCam 
proved that Lucky Imaging is a very promising alternative to adaptive optics, allowing diffraction 
limited imaging at telescopes in the 2--3\,m class at wavelengths below 1$\mu$m \cite{Baldwin:2001,Tubbs:2002,Tubbs:2003,Law:2006}.

These results triggered the development of a similar instrument for the 2.2-m telescope at
Calar Alto by our group. 
The Lucky Imaging camera AstraLux was built in less than 5 months thanks to the availability of most 
parts as off-the-shelf equipment. 

In the following we describe the instrument design and data processing pipeline and
summarise the instrument's performance and key characteristics. Interested readers
looking for more comprehensive information are referred to the diploma 
thesis, which covers AstraLux and its performance in full detail \cite{Hormuth:2007a}\footnote{Available online at \texttt{http://www.mpia.de/ASTRALUX/}}.

\section{Instrument Design}

\begin{figure}
	\begin{center}
   		\begin{tabular}{c}
			\includegraphics[width=13cm]{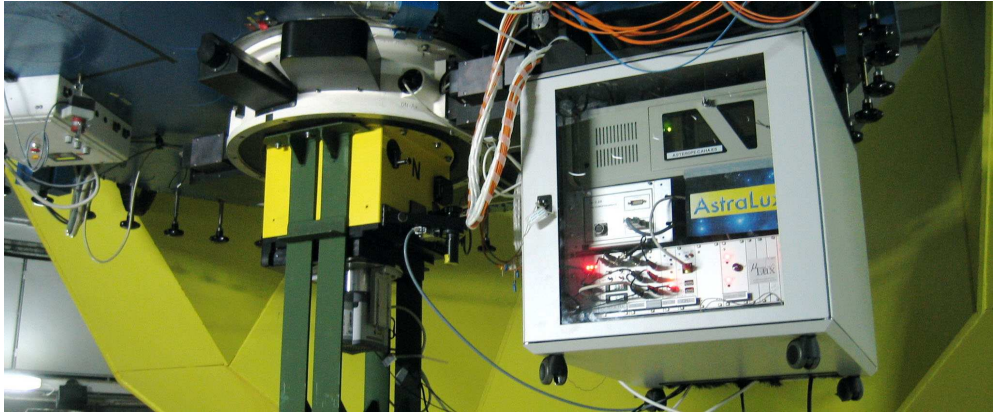}
 	 	 \end{tabular}
 	  \end{center}
  	\caption[example] 
		{ \label{fig:telescope} 
	         AstraLux at the Calar Alto 2.2-m telescope
		telescope. The yellow box contains the 8-position filterwheel, the camera is attached at the bottom.
		The grey rack to the right houses the camera control computer, a keyboard/monitor combination,
		the filterwheel electronics, and MicroLux, a GPS timing add-on.}
\end{figure} 
\subsection{Telescope and Site Characteristics}
The Calar Alto observatory is located at a height of 2150\,m in the Sierra de los 
Filabres, approximately 50\,km north of Almeria in Andalusia, southern Spain. 
The median seeing is 0\farcs9, slightly better in summer than in winter \cite{Sanchez:2007}.

Together with the given telescope diameter of 2.2\,m, the seeing defines the optimal wavelength for
Lucky Imaging: 
the probability for the occurence of a good frame (i.e. with a Strehl ratio $>$37\%) in 
a series of short exposure time images decreases exponentially \cite{Fried:1965} with the ratio of 
telescope diameter over the Fried parameter $r_0$, which again depends inversely on the seeing 
and scales with $\lambda^{5/6}$. 
While observations at short wavelengths result in too low probabilities for usable frames,  
the gain in resolution by selecting only high-quality frames becomes small at long wavelengths 
due to the decreasing theoretical resolution of the telescope. 
For the Calar Alto 2.2-m telescope, the optimal wavelength for efficient Lucky Imaging lies in the range 
of 800--1000\,nm, including the Johnson $I$ and the Sloan Digital Sky Survey (SDSS) $i'$ and $z'$ 
passbands. At typical seeing conditions one can expect that $\approx$0.5--1\% of all images will have 
an acceptable Strehl ratio of few 10\% while providing a 6 to 10 times better resolution.

\subsection{Camera \& Optics}
The AstraLux camera head is an Andor DV887-UVB model from Andor Technologies,
Belfast, Northern Ireland. It is an electron-multiplying, thinned, and back-illuminated 512$\times$512 
pixel CCD that comes as complete package with a multi-stage Peltier cooler, mechanical shutter, computer 
interface, and software. 

Starting at room temperature, the typical operating temperature of $-$75\degr C is reached within less than 10\,minutes. 
The camera requires neither refill of liquid coolants nor any action to maintain the vacuum inside the CCD head.
The quantum efficiency (QE) of the E2V CCD97 detector peaks at $>$90\% in the $R$ band. At 912\,nm,
the central wavelength of the SDSS $z'$ filter, the QE is still $\approx$40\%.
For Lucky Imaging we use a readout clock of 10\,MHz, giving a frame rate of $\approx$34\,Hz at an A/D resolution of 14\,Bit.
The readout noise is $\approx$80e$^-$.
At an electron gain of 2500, this corresponds to a SNR of $>$30 for single photons. 

At the Calar Alto 2.2\,m telescope, the camera's physical pixel size of 16\,$\mu$m roughly  corresponds
 to almost twice the size of the theoretical PSF at 912\,nm (SDSS $z'$ band). 
For the final design, a pixel scale of 47\,mas/px was adopted as a good compromise between spatial sampling 
and the resulting size of the field of view which is 24$\times$24\arcsec\ .
The hardware realisation chosen for the AstraLux instrument is a single negative achromat in
Barlow configuration, i.e. placed in the optical path before the nominal focal plane of the telescope.

The instrument's filter wheel with 8 positions can hold
virtually any filter that is available at the observatory\footnote{See \texttt{http://www.caha.es//CAHA/Instruments/filterlist.html}
for a list of all available filters.}, allowing observations at a wide range of wavelengths. 
The standard filters used for Lucky Imaging observations are SDSS $i'$ and $z'$ band interference
filters, manufactured by Asahi Spectra Co., Ltd., Tokyo, Japan. The transmission curves are very
close to the original SDSS filter system definition \cite{Fukugita:1996} and peak at more than 95\%.

\subsection{Instrument Control \& Data Processing}
AstraLux can be controlled from virtually any point of the observatory with a 100\,MBit Ethernet
network connection. 
The camera control software offers a real time display and allows setting of all crucial parameters 
like exposure time, electron gain, or number of requested frames. The raw FITS data cubes (typically 18\,MB/s
at maximum frame rate) are not stored on the camera computer itself, but on a fast RAID-0 array
with 1\,TB capacity in a gateway machine.

The AstraLux  data reduction software is running on a dedicated
pipeline computer, equipped with two dual-core Woodcrest processors and 8\,GB of memory.
The pipeline automatically produces quicklook results of the Lucky Imaging observations 
in approximately the same time that is needed for data acquisition. The basic pipeline algorithms are 
similar to that adopted by the LuckyCam team \cite{Tubbs:2003a,Law:2006a}.

After completion of a Lucky Imaging observation, the position of a suitable reference 
object for quality assessment has to be determined. 
This is performed on a stacked image  of the first 2 seconds of raw data, and can be 
done either manually or automatically. 
The pipeline tracks the reference on all following raw data frames to account
for large atmospheric tip/tilt or telescope tracking errors.
Subsequently, the quality of each single frame is determined by measuring the Strehl ratio of
the reference object after extracting, resampling, and noise-filtering a small region around the
reference. 

The pipeline's image reconstruction module performs data reduction in its literal sense. From
typically several GB of input data, just a few MB of pipeline results are produced. The most
interesting ones -- the Lucky Imaging results -- are currently generated with the Drizzle
algorithm \cite{Fruchter:2002}. This linear reconstruction method is flux preserving
and able to at least partially overcome the slight undersampling that is present in the raw data. 
It is capable of handling sub-pixel translations without the need to perform image shifting in the Fourier
domain. 

The drizzling process oversamples the input data twice, resulting in a pixel scale of $\approx$23.7\,mas/px in the final images.
Currently the pipeline produces drizzled results of the best 1, 2.5, 5, and 10\% of the input frames.
Bias and flatfield calibration frames are applied to the input images prior to drizzling.
A seeing (and tracking) limited image  is generated as well using all frames to allow quick measurements 
of the seeing conditions, useful at times when the observatory's seeing monitor is switched off, or for later
assessment of the data quality.

\begin{figure}
\begin{minipage}{8cm}
\includegraphics[height=8cm,angle=270]{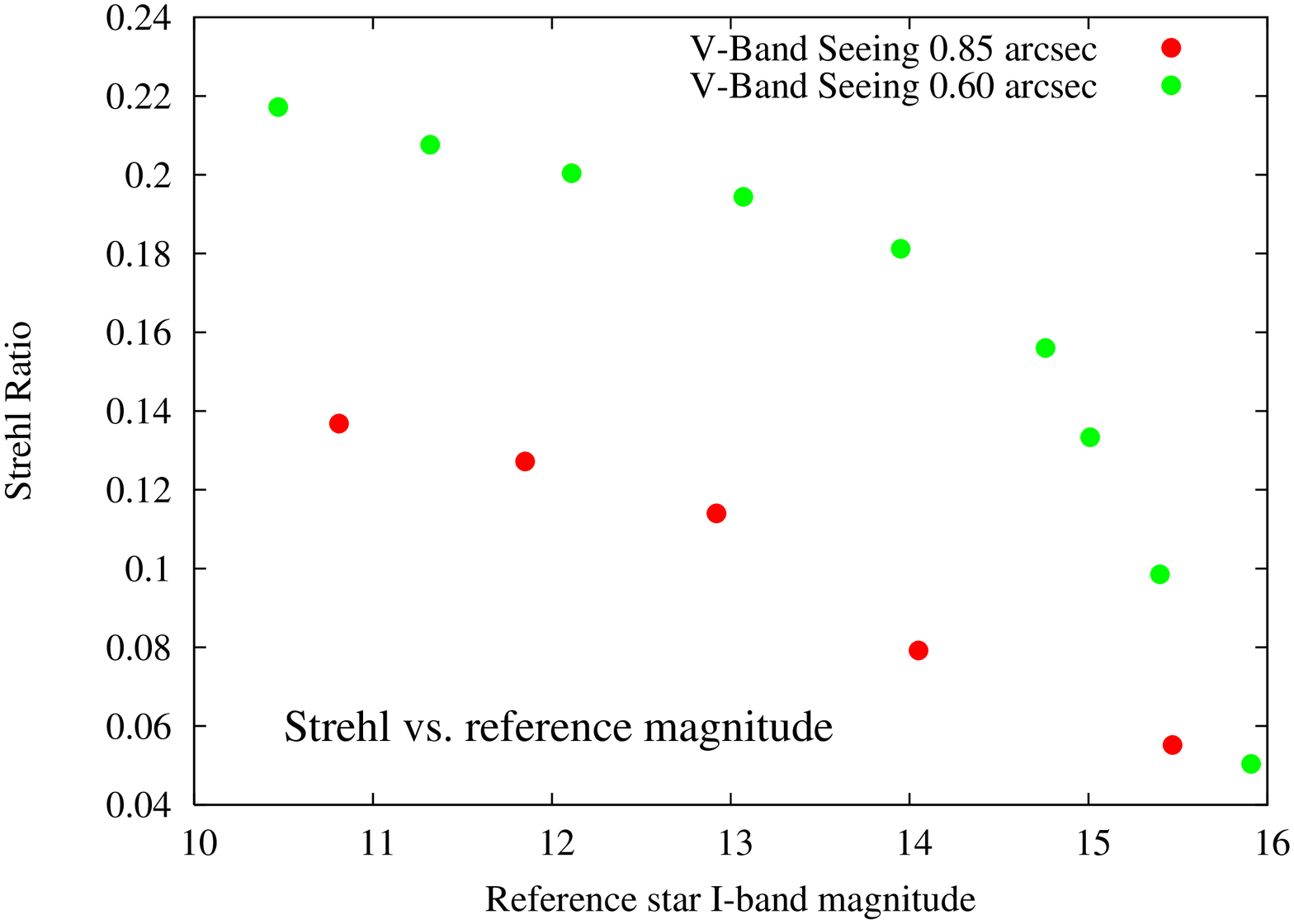}
\caption{\label{fig:strehlvsmag}Dependency of the final Strehl ratio on natural seeing and reference star magnitude.
		The plot is based on observations with 30\,ms single frame exposure time in the SDSS {\em z'}		
		filter and 1\% image selection rate.}
\end{minipage}\hspace{2pc}
\begin{minipage}{7cm}\vspace{-8.2mm}
\includegraphics[height=8.3cm,angle=270]{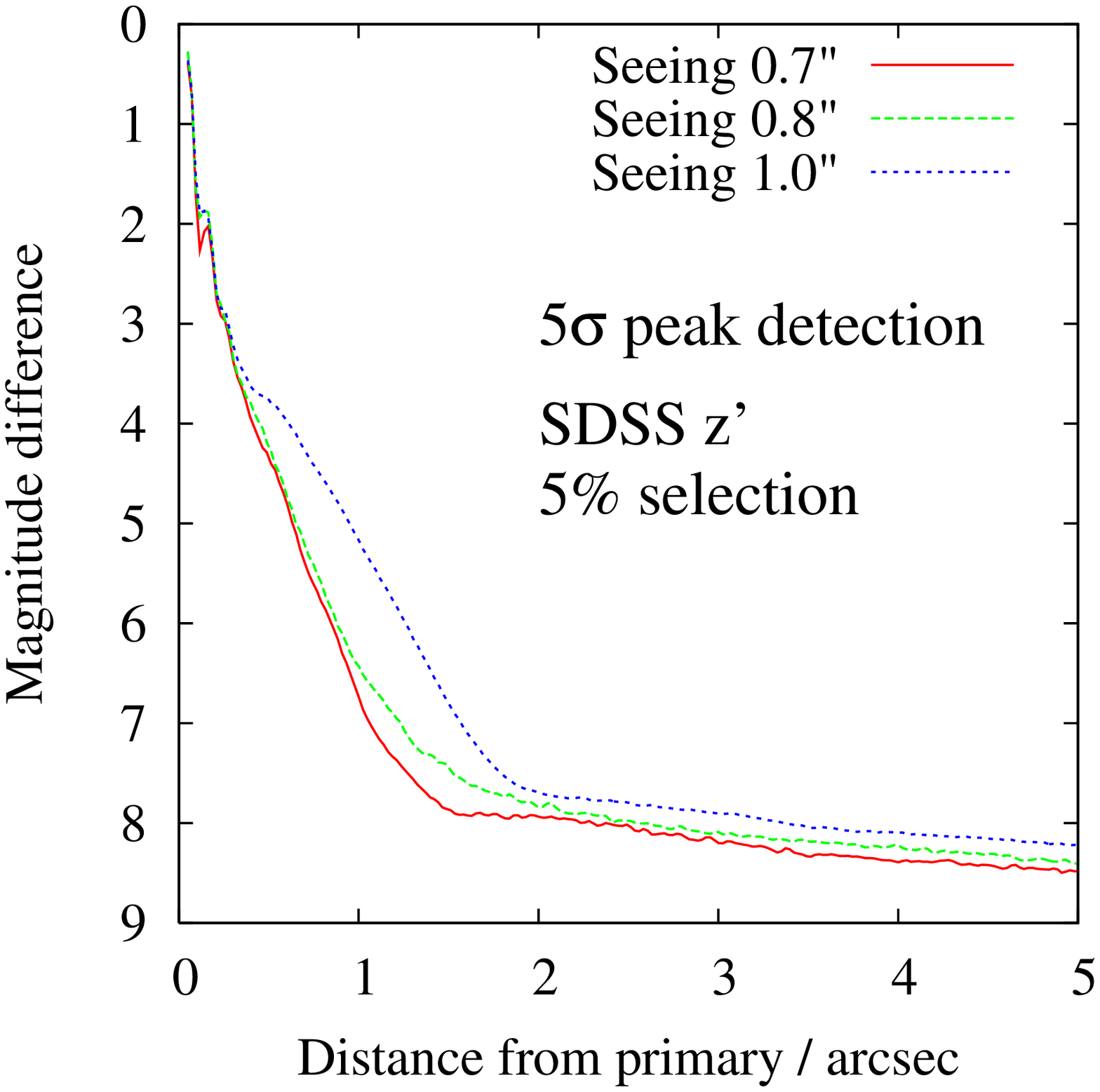}
\caption{\label{fig:magdiff}Typical achievable magnitude differences for a 5$\sigma$ peak detection of a fainter companion
		to the reference star.}
\end{minipage} 
\end{figure}

\section{System Performance}
\begin{figure}
\includegraphics[width=8.1cm]{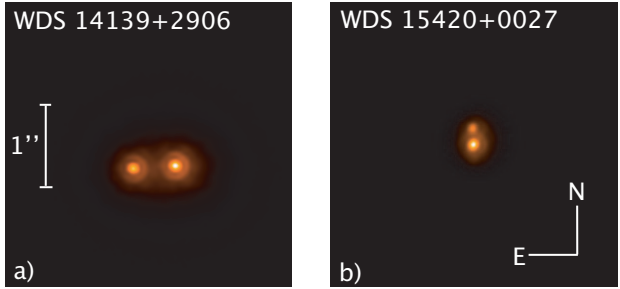}\hspace{2pc}%
\begin{minipage}[b]{7cm}\caption{\label{fig:first22}AstraLux first light: the binary WDS\,14139+2906 with an angular
		separation of 0\farcs52 and  magnitudes {\em V}=7.5 and {\em V}=7.6\,mag, and 
		the 0\farcs23 separated binary WDS\,15420+0027 with {\em V}=8.2 and {\em V}=8.8\,mag
		 brightnesses. Both images are based on a 2\%-selection from 10000 single frames
		in the SDSS {\em z'} filter with 30\,ms exposure time each. Image scaling is linear.}
\end{minipage}
\end{figure}
First light at the Calar Alto 2.2-m telescope was obtained on July 6, 2006 with 
photometric sky conditions and {\em V}-band seeing values as low as 0\farcs6.

Known bright double stars were chosen as first light targets and observed in the SDSS {\em z'}\ 
band with resulting Strehl ratios of $\approx$20\% (see Figure\,\ref{fig:first22}).
The point spread function was found to show a remarkable radial symmetry, with a typical FWHM
of the PSF core of 110\,mas in the SDSS {\em z'} filter. 

Operating the instrument, and especially acquiring targets, proved to be much easier than anticipated.
Though the pointing accuracy of the telescope is in general not better than 10\arcsec, the availability of
the camera's real time display allows short acquisition times of typically 1--2\,min per target.

\begin{figure}
	\begin{center}
   		\begin{tabular}{c}			
   			\includegraphics[width=12.2cm]{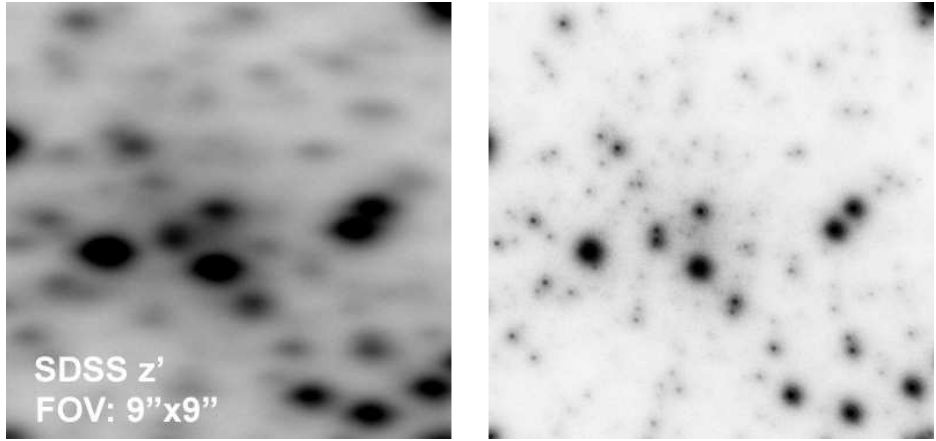}
 	 	 \end{tabular}
 	  \end{center}
  	\caption[example] 
		{ \label{fig:M15} 
	 Comparison between seeing limited imaging and the ``Lucky'' version: The combination of the best 5\% of 10000 single frames provided a Strehl ratio of 20\% in this observation of the core of the globular cluster M15. Though the conventional result contains 20 times more photons, it is clearly inferior in terms of point source detection limits. 
	}
\end{figure}

AstraLux observations of globular cluster centres enabled the characterisation of the image quality
over the full field of view. Choosing different stars with a wide
range of magnitudes as the Lucky Imaging reference allowed to estimate brightness limits
for the reference selection and to measure the dependency of the Strehl ratio on the reference magnitude.
Among the globular clusters M3, M13, and M15, the latter  has been observed most extensively
with AstraLux. Figure~\ref{fig:M15} shows a comparison of a Lucky Imaging result to the corresponding
seeing limited image.

\subsection{Isoplanatic Angle, Coherence Time \& Limiting Magnitudes}
The observations of centres of the globular clusters M13 and M15 allowed measurements of the Strehl ratio for
a large number of stars, well distributed over the field of view. The values were normalised by the
Strehl ratio of the reference star, and the isoplanatic angle was determined by finding the angular separation  
from the reference where the Strehl ratio drops to $1/e$.
This procedure was applied to observations with different exposure times ranging from 15 to 60\,ms, 
consistently resulting in an estimate of the isoplanatic angle of $\approx$40\arcsec\ in the SDSS {\em z'} band.

To determine the speckle pattern coherence time, a series of 10000 images of the bright star $\beta$\,{\em And} 
was recorded with a time resolution of 4.6\,ms. The measured coherence time $\tau_e$=36\,ms corresponds to the
1/$e$ point of a Lorentzian fit to the temporal autocorrelation function of the focal plane intensity at a fixed point.

The globular cluster observations were re-analysed with different choices of the Lucky Imaging
reference star to assess the impact of the reference magnitude on the final Strehl ratio.
Figure~\ref{fig:strehlvsmag} shows the results for measurements under two different seeing conditions.
While reference stars as faint as $I$=15.5\,mag still allow a substantial improvement of image
quality under a 0\farcs65 seeing, the same performance cannot be reached with stars fainter
than 13.5\,mag in 0\farcs85 seeing. 

The achievable contrast ratio for close companions to bright host stars was determined on 
final pipeline results of SDSS\,$z'$ band observations of single stars under different seeing conditions. 
The deduced maximum magnitude differences for a 5$\sigma$ peak detection are based on 
measurements of the noise in concentric rings around these stars. 
Simulations with observed PSFs were carried out to check the reliability of the numerical results. 
Figure~\ref{fig:magdiff} shows typical detection limit plots for three different $V$-band seeing values. 
At angular separations larger than 2\arcsec, the detection limit is determined by readout noise and 
the Poisson noise of the sky background.


\section{Conclusions}
Within less than one year it was possible to design, build, and characterise a Lucky Imaging
instrument for the Calar Alto 2.2-m telescope. As a common user instrument from 2007 on, AstraLux has become the standard 
tool for diffraction limited imaging at the Calar Alto observatory. It is currently mostly used for
binarity surveys among stars and minor planets. 

AstraLux is able to reach Strehl ratios as high as 25\% in the {\em z'} band. 
In general, Lucky Imaging provides an improvement of the Strehl ratio by a 
factor of 10, corresponding to an increase of the signal-to-noise ratio for point sources by
a factor of 10$-$20, depending on atmospheric conditions.
Thus a selection of only the best 5$-$10\% of all images does definitely not have a negative effect on
the detection limit for point sources.

The requirements for the reference star magnitude are similar as for observations with 
adaptive optics.
The performance starts to significantly decrease at $I$=14\,mag, but 
image quality improvements are still possible with stars as faint as 15$-$16\,mag. 
The measured isoplanatic angle in the $z'$-band is with $\approx$40\arcsec\ as large as 
typical values in $K$-band for adaptive optics observations. 

The measured close companion detection limit at an angular separation of 1\arcsec\ is
on average 6\,mag, worse than what adaptive optics can provide.
But: adaptive optics at 8-m class telescopes currently has this capability only in the $H$ and $K$-band at 
wavelengths $>$1.5\,$\mu$m.
The achievable contrast ratio in speckle imaging observations is typically two 
magnitudes less than for Lucky Imaging.

The encouraging results regarding both development speed as well as scientific output
triggered the start of the project ``AstraLux Sur'', an almost identical copy of the Calar Alto
version to be used as visitor instrument at ESO's New Technology Telescope (NTT) at La Silla, Chile.
AstraLux Sur is currently on its way to Chile after a development time of less than 2 months and 
will have first light in mid-July 2008.

\section*{References}
\bibliography{astralux}   
\bibliographystyle{iopart-num}   

\end{document}